\documentstyle[12pt,a4]{article}

\newcommand{\mb}[1]{\ifmmode#1\else\mbox{$#1$}\fi}
\newcommand{\as}{\mb{\alpha_s}}
\newcommand{\ee}{\mb{\mathrm{e^+e^-}}}

\newcommand{\dP}{\mathrm{d}{\cal{P}}}
\newcommand{\incl}{_{\mathrm{incl}}}
\newcommand{\first}{_{\mathrm{1st}}}
\newcommand{\ps}{^{\mathrm{p.s.}}}
\newcommand{\me}{^{\mathrm{m.e.}}}
\newcommand{\hard}{_{\mathrm{hard}}}
\newcommand{\soft}{_{\mathrm{soft}}}
\newcommand{\xp}[1]{\exp\biggl\{{#1}\biggr\}}
\newcommand{\xmax}{{x_{\mathrm{max}}}}
\font\bigm=cmex10 scaled\magstep1
\def\bigint_#1^#2{\hbox{\raise16.8pt\hbox{\bigm\char'132}}_{#1}^{\;\;#2}}

\input FEYNMAN 
\gdef\drawfermion[#1#2](#3,#4)[#5,#6][#7]{ 
  \newcount\Xp
  \newcount\Yp
  \newcount\Xd
  \newcount\Yd
  \newcount\Xs
  \newcount\Ys
  \newcount\Xl
  \newcount\Yl
  \gdef\drawfermion[##1##2](##3,##4)[##5,##6][##7]{
    \Xp=##3 \Yp=##4 \Xd=##5 \Yd=##6 \Xl=##7
    \multiply\Xd by \Xd
    \multiply\Yd by \Yd
    \Xs=\Xd
    \advance\Xs by \Yd
    \Xd=##5 \Yd=##6 \Ys=500000
    \ifnum \Xs=1 \Xs=1000      
    \else\ifnum \Xs=2 \Xs=1414 
    \else\ifnum \Xs=5 \Xs=2236 
    \else\ifnum \Xs=10\Xs=3162 
    \else\ifnum \Xs=17\Xs=4123 
    \else\ifnum \Xs=13\Xs=3606 
    \else\ifnum \Xs=25\Xs=5000 
    \else\Xs=1000
    \fi\fi\fi\fi\fi\fi\fi
    \divide\Ys by \Xs
    \Xs=\Xl
    \multiply\Xs by \Ys \divide\Xs by 1000
    \Ys=\Xs
    \multiply\Ys by \Yd
    \multiply\Xs by \Xd
    \ifnum \Xs=0 \Xl=\Ys \else \Xl=\Xs \fi
    \ifnum \Xl<0 \negate\Xl \fi
    \ifdim##2=\ATTIP
      \Yl=0
    \else
      \Yl=\arrowlength
      \multiply\Yl by \Xl
      \multiply\Yl by 2
      \divide\Yl by ##7
    \fi
    \global\pfrontx=\Xp \global\pfronty=\Yp
    \put(\Xp,\Yp){\line(\Xd,\Yd){\Xl}}
    \advance\Xp by \Xs \advance\Yp by \Ys
    \global\pmidx=\Xp \global\pmidy=\Yp
    \put(\Xp,\Yp){\line(\Xd,\Yd){\Xl}}
    \ifnum##1=\REG
      \put(\Xp,\Yp){\vector(\Xd,\Yd){\Yl}}
    \else
      \put(\Xp,\Yp){\vector(-\Xd,-\Yd){\Yl}}
    \fi
    \advance\Xp by \Xs \advance\Yp by \Ys
    \global\pbackx=\Xp \global\pbacky=\Yp
  }
  \drawfermion[#1#2](#3,#4)[#5,#6][#7]
}
\Feynmanlength  

\catcode`\@=11
\newlength{\captionsize}
\newlength{\captionlength}
\long\def\@caption#1[#2]#3{\par\addcontentsline{\csname
  ext@#1\endcsname}{#1}{\protect\numberline{\csname
  the#1\endcsname}{\ignorespaces #2}}\begingroup
    \@parboxrestore
    \normalsize
    \@makecaption{\csname fnum@#1\endcsname}{\ignorespaces
\settowidth{\captionsize}{#1~\csname the#1\endcsname:}%
\setlength{\captionsize}{-\captionsize}%
\addtolength{\captionsize}{\textwidth}%
\addtolength{\captionsize}{-3mm}%
\settowidth{\captionlength}{#3}%
\ifdim\captionlength<\captionsize{#3}\else%
\parbox[t]{\captionsize}{#3}%
\fi%
}\par
  \endgroup}
\catcode`\@=12

\begin{document}
\begin{titlepage}
\addtocounter{page}{1}
\begin{flushright}
     hep-ph/9410414 \\
     LU TP 94-17 \\
     October 1994
\end{flushright}
\par \vskip 15mm
\vspace*{\fill}
\begin{center}
\Large\bf\boldmath
                Matrix-Element Corrections to \\
                  Parton Shower Algorithms
\end{center}
\par \vskip 2mm
\begin{center}
        {\bf Michael H.\ Seymour}%
\footnote{Address after 1st January 1995: CERN TH Division, CH-1211
Geneva 23, Switzerland.} \\
        Department of Theoretical Physics, University of Lund, \\
        S\"olvegatan 14A, S-22362 Lund, Sweden
\end{center}
\par \vskip 2mm

\begin{center} {\large \bf Abstract} \end{center}
\begin{quote}
\pretolerance 1000 
We discuss two ways in which parton shower algorithms can be
supplemented by matrix-element corrections to ensure the correct hard
limit: by using complementary phase-space regions, or by modifying the
shower itself.  In the former case, existing algorithms are
self-consistent only if the total correction is small.  In the latter
case, existing algorithms are never self-consistent, a problem that is
particularly severe for angular-ordered parton shower algorithms.  We
show how to construct self-consistent algorithms in both cases.
\end{quote}
\vspace*{\fill}
\begin{flushleft}
     LU TP 94-17 \\
     October 1994
\end{flushleft}
\end{titlepage}

\section{Introduction}
Monte Carlo event generators[\ref{HW}--\ref{AR}] that combine the parton
shower[\ref{MW},\ref{Sj}] or dipole cascade[\ref{GP}] approaches for
perturbative jet evolution with local models of non-perturbative
hadronisation[\ref{W},\ref{AGIS}] have been extremely successful in
describing the hadronic final state of high energy reactions (see
[\ref{rev}] for example).  One feature that is essential for an accurate
description of final state effects is the coherence of radiation from
different partons in the event.  In the parton shower approach, this can
be implemented as an ordering in opening angles during the evolution
away from the hard process.  In the virtuality-ordered algorithm
of~[\ref{Sj}], this is done by vetoing non-ordered emission, while in
the algorithm of~[\ref{MW}], it is more directly incorporated by using
opening angle as the evolution variable.

However, because parton showers are based on expansions around the soft
and collinear limits that dominate the total emission cross-section,
there is no guarantee that they will perform well away from those
limits.  Many experimental observables are specifically sensitive to
hard emission, such as event shapes in the 3-jet region of \ee\
annihilation, or the $E_T$ flow in DIS.  To give a reliable prediction
for such quantities, it is necessary to supplement the parton shower
algorithm with the exact first-order matrix-element cross-section.  This
is particularly important for angular-ordered algorithms.

There are two approaches in current practice: either to split the
phase-space into two parts, using the matrix-element cross-section in
one region and the parton shower in the other[\ref{MEPS}]; or to modify
the algorithm so that it faithfully reproduces the matrix-element
cross-section in the hard limit[\ref{Sj}].  It is worth noting that the
dipole cascade model[\ref{GP}] automatically reproduces this hard limit
by construction, so no such matrix-element corrections are required.

In this paper, we discuss both approaches and show that the former can
be inconsistent if the total correction is large.  In the latter,
particular care must be taken to ensure the theoretical consistency.  We
show how self-consistency can be achieved in both cases, with minor
modifications to existing algorithms.

\section{Basics}
We begin by recalling some of the features of parton shower evolution.
We consider an arbitrary hard process with cross-section $\sigma_0,$ and
study the cross-section for additional radiation,
\begin{equation}
  \dP\incl\me = \frac1{\sigma_0}\mathrm{d}\sigma.
\end{equation}
The subscript refers to the fact that the matrix-element cross-section
describes the {\em inclusive} emission rate.  After integrating over the
whole of phase-space, this gives the average number of emissions,
\begin{equation}
  N\incl = \int\dP\incl\me.
\end{equation}
Unless an infrared cutoff is applied, $N\incl$ is a divergent quantity.

Parton shower algorithms are based on a `sequential evolution' picture
in which multiple emission occurs in a definite (but algorithm-specific)
order, for example from largest to smallest angle in~[\ref{MW}] or
largest to smallest virtuality in~[\ref{Sj}].  This is defined by some
{\em ordering variable} $q^2\footnote{This is an arbitrary function of
the emission kinematics, not necessarily the pair virtuality.},$ with
emissions with larger $q^2$ being treated earlier in the evolution than
those with smaller $q^2$.  An important quantity for the construction of
the algorithm is the probability that there was no emission before some
scale $q^2$ (which is generally referred to as a form factor, although
it is formally the ratio of two form factors),
\begin{equation}
  \Delta(q^2) = 1-\int_{q^2}\dP\incl\me +
    {\cal{O}}\left({\textstyle{\frac1{2!}}}
      \int_{q^2}{\dP\incl\me}^2\right) -
    \ldots
  \approx \xp{-\int_{q^2}\dP\incl\me},
\end{equation}
where the approximation is valid when the cross-section is dominated by
the soft and collinear regions, i.e.~when $q^2$ is small.  Note that we
do not explicitly give the upper limit of the integral because it
depends on the definition of $q^2$.  One of the approximations used to
construct a probabilistic parton shower algorithm is to replace the
approximation by an equality,
\begin{equation}
  \Delta(q^2)
  \equiv \xp{-\int_{q^2}\dP\incl\me}.
\end{equation}
The cross-section for the {\em first} emission is then simply the
product of the inclusive cross-section with the probability that there
was no earlier emission,
\begin{equation}
  \dP\first\me(q^2) = \dP\incl\me(q^2)\xp{-\int_{q^2}\dP\incl\me},
\end{equation}
which is always finite.  One then obtains the average number of first
emissions,
\begin{equation}
 N\first = \int\dP\first\me \le 1,
\end{equation}
where the equality applies when there is no infrared cutoff.  The other
approximation used is to replace the exact cross-section by
\begin{eqnarray}
  \dP\incl\ps &=& \frac1{\sigma_0}\mathrm{d}\sigma\ps, \\
  \dP\first\ps(q^2) &=& \dP\incl\ps(q^2)\xp{-\int_{q^2}\dP\incl\ps},
\end{eqnarray}
which is again valid in the soft and collinear limits.  In general for
multiple emission, only this approximation can be calculated, and not
the exact matrix element.  This paper is about how the exact
cross-section can be used to improve the parton shower algorithm in
cases in which it is known.  Specifically, we consider the first-order
matrix element, which dominates the cross-section when there is one
emission that is much harder than all others.

\section{Complementary phase-spaces}
Since the first-order matrix-element cross-section is reliable in the
hard limit, and parton shower algorithms are reliable in the soft and
collinear limits, the simplest approach is to split the phase-space into
two parts, and use each in their reliable regions.  If, as one would
hope, there is reasonable overlap between the two regions of
reliability, then the final result should not depend strongly on where
the border between the two regions is drawn.  To prevent double-counting
of emissions, the border must be used consistently in both the
matrix-element and parton shower phase spaces, i.e.~the two regions must
be exactly complementary.

In the model of~[\ref{MEPS}], this is implemented for DIS by using an
adjustable cutoff in invariant mass.  On the other hand, using the
angular-ordered parton shower algorithm of~[\ref{MW}] one finds that the
shower phase space has a natural border, from the requirement that all
emissions be into the forward hemisphere in the particular Lorentz frame
used.  This was exploited in~[\ref{S1}] for \ee\ and~[\ref{S2}] for DIS,
to provide a similar correction, but without an adjustable cutoff.

Since the first-order matrix element describes the {\em inclusive}
emission cross-section, it only reliably predicts the {\em
single-emission} cross-section if the region in which it is used
contributes a small fraction of the total cross-section.  If this
fraction approaches unity, then multiple emission is bound to play a
r\^ole and the matrix-element cross-section will become unreliable.  In
the \ee\ algorithm of~[\ref{S1}], this fraction is a very safe 1~in~40,
while for DIS[\ref{S2}] it is typically 1~in~10 so still reasonably
safe.  On the other hand, the cutoff in~[\ref{MEPS}] is
phenomenologically required to be rather small, leading to a fraction
that is typically around 50\%.  They also find a residual dependence on
the cutoff value.  Thus we should consider how the way in which the
matrix-element is used might be modified to improve the agreement.

Although we have said that the matrix-element cross-section only
generates a single emission in its phase-space region, multiple emission
is in fact generated by the algorithm.  This is because after generating
an emission within that region, the resulting hard process is parton
showered with upper limits for emission controlled by the dynamics of
the hard emission.  In particular, subsequent emission from this system
cannot be harder than the hard emission, but can be harder than the
cutoff between the two regions.  Thus, to the same accuracy as the usual
parton shower, multiple emission into the matrix-element region is
generated.

Since the parton shower emission cannot be harder than the emission
generated using the matrix element, the matrix-element emission must be
the hardest in the event.  However, this is in contradiction with the
fact that the matrix element describes the inclusive emission
cross-section.  To avoid this contradiction, a form factor should be
included to incorporate the probability that there was not an emission
at a higher scale.  That is, instead of generating events in the
matrix-element region according to the inclusive cross-section,
$\dP\incl\me,$ they should be generated according to $\dP\first\me,$
which correctly accounts for the probability that there was no emission
at a higher scale than the one generated.

\begin{figure}
  \vspace*{5.3cm}
  \includegraphics{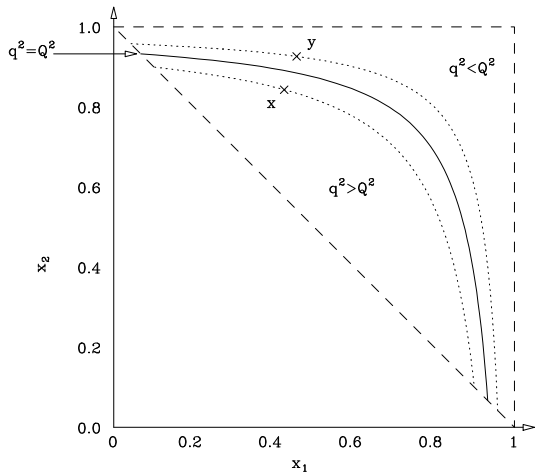}
  \caption{The phase-space for \ee\ divided according to an ordering
    variable $q^2$ into a matrix-element region, $q^2>Q^2,$ and a parton
    shower region, $q^2<Q^2$.  Recall that the matrix element is
    divergent at $x_{1,2}=1$.}
  \label{ee}
\end{figure}

In figure~\ref{ee} we show an illustration for the specific case of \ee\
annihilation.
The probability distribution for the first emission to be at y is
\begin{equation}
  \dP\first\ps(q_y^2) = \dP\incl\ps(q_y^2)
    \xp{-\int_{q_y^2}^{Q^2}\dP\incl\ps}
    \xp{-\int_{Q^2}        \dP\incl\me}.
\end{equation}
The second integration arises because the first emission can only come
from the parton shower if there was no matrix-element emission.  If
events at x in the matrix-element region are generated according to the
exact first-order matrix element, as in the standard algorithm, they are
distributed according to $\dP\incl\me(q_x^2)$.  In the limit
$q_{x,y}^2\to Q^2,$ where the points correspond to identical physical
configurations, the two probability distributions are different, even if
$\dP\incl\ps$ is a perfect approximation to $\dP\incl\me,$ leading to a
residual dependence on the cutoff between the two regions,~$Q^2$.  If
\begin{equation}
  \int_{Q^2} \dP\incl\me
\end{equation}
is large, this dependence is strong.  On the other hand, if we generate
the matrix-element events according to
\begin{equation}
  \dP\first\me(q_x^2) = \dP\incl\me(q_x^2)\xp{-\int_{q_x^2}\dP\incl\me},
\end{equation}
as we propose, then in the $q_{x,y}^2\to Q^2$ limit we obtain
\begin{eqnarray}
  \dP\first\me(Q^2) &=& \dP\incl\me(Q^2)\xp{-\int_{Q^2}\dP\incl\me}, \\
  \dP\first\ps(Q^2) &=& \dP\incl\ps(Q^2)\xp{-\int_{Q^2}\dP\incl\me}.
\end{eqnarray}
If the parton shower cross-section exactly reproduced the first-order
matrix element, the two would then be perfectly matched, with no
dependence on $Q^2$.

\section{Correcting the algorithm}
In~[\ref{Sj}], a method was described to correct the first emission of a
virtuality-ordered algorithm to reproduce the first-order matrix-element
cross-section.  Although we shall show that this is not self-consistent,
it is similar to the method we propose, so it is worth describing the
details.

If one stops a parton shower after one emission then the final state is
exactly that described by the first-order matrix element: three partons
in the \ee\ case.  It is straightforward to then work through the
kinematic reconstruction used by the algorithm, to relate the variables
generated in the parton shower branching to those used to describe the
matrix-element cross-section, and calculate the differential
cross-section produced by the algorithm, $\dP\incl\ps$.  In the
algorithm of~[\ref{Sj}], this was found to be everywhere larger than the
matrix-element cross-section, $\dP\incl\me,$ so the veto algorithm was
used to correct the distribution of first branchings\footnote{See the
appendix for a brief discussion of the veto algorithm.}.

However, we recall that the time-ordered language used to describe
parton shower evolution is not fundamental to the theory, so when we use
a concept like the `first' emission we must be extremely careful that we
have retained theoretical consistency.  Indeed, the important point has
been made above, that the first-order matrix-element cross-section
describes the inclusive distribution of all emissions from the original
current, and not just the first.  Thus {\em all} hard emission should be
corrected to the first-order matrix-element cross-section, and not just
the first.

At this point it becomes necessary to define more specifically what we
mean by `hard' emission.  The most suitable definition is in terms of
transverse momentum, since this avoids in a single variable both the
soft and collinear singularities.  By contrast, the virtuality and
opening angle, which are used as ordering variables in the algorithms
of~[\ref{Sj}] and~[\ref{MW}] respectively, do not prevent emission at
low transverse momenta where the running coupling becomes large.  To
prevent them from entering the non-perturbative region, the infrared
cutoff is active throughout the evolution.  Furthermore, the QCD matrix
elements for multiple emission factorise in the limit of
strongly-ordered transverse momenta, so using it as the variable to
measure hardness allows simple construction of algorithms.

In terms of the transverse momentum, the recoil of parton $a$ from the
emission of parton $b$ is extremely simple---parton $a$'s direction is
perturbed by an amount proportional to the transverse momentum.  Thus we
can simply analyse the effect of a later emission from parton $a$ in the
two strongly-ordered regions, in which the second emission is much
harder or much softer than the first.  In the first case, as far as the
second emission is concerned, the recoil from the first emission is
insignificant, so the emitting current is identical to the original
current, and the emission should be described the first-order matrix
element.  Furthermore, since the first emission can be considered
infinitely soft by comparison, the second emission can be related to the
matrix-element cross-section by pretending that the first emission never
occurred, i.e.~exactly as in the method described above.  On the other
hand if the second emission is much softer than the first, it is
effectively emitted by a completely different current and should not be
corrected to the first-order matrix-element cross-section.  Instead, the
second-order matrix element factorises in this limit, and the parton
shower algorithm is reliable without correction.

Thus, we see that the parton shower algorithm can be corrected to the
first-order matrix-element cross-section by applying the method given
above to {\em every emission that is the hardest so far,} instead of
just the first.

\begin{figure}
\begin{center}
\begin{picture}(22000,10000)
\pbackx=11000\pbacky=5000
\THICKLINES
\bigphotons
\global\advance\pbackx by -10250
\global\advance\pbacky by  -1900
\put(\pbackx,\pbacky){\boldmath$\times$}
\global\advance\pbackx by 500
\global\advance\pbacky by 300
\drawfermion[\REG\ATBASE](\pbackx,\pbacky)[1,0][5000]
\drawline\gluon[\N\REG](\pbackx,\pbacky)[3]
\drawfermion[\REG\ATBASE](\pfrontx,\pfronty)[1,0][5000]
\drawline\gluon[\NE\REG](\pbackx,\pbacky)[5]
\drawfermion[\REG\ATBASE](\pfrontx,\pfronty)[4,-1][10000]
\end{picture}
\end{center}
\vspace*{-4ex}
  \caption{Emission of two gluons in which the first (according to the
    ordering variable of the algorithm) is softer, and at a larger
    angle, than the second.  Both should be described by the first-order
    matrix element: the second because the recoil from the first is
    negligible,
    so it is effectively emitted by the original current; and
    the first because it represents the coherent sum of emissions from
    the external lines which, after azimuthal averaging, is equivalent
    to a single emission from the internal line pretending that the
    later emission did not happen, i.e.~it is also effectively emitted
    by the original current.}
  \label{soft}
\end{figure}
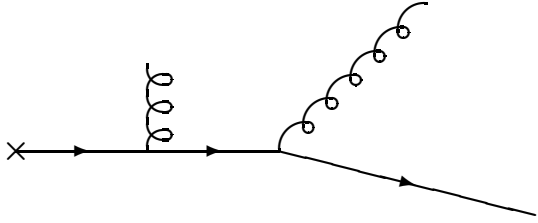

One might suppose that having found a later emission that was harder
than the first, the first should be considered as coming from a modified
current, so should not have been corrected.  This is not the case, owing
to the coherence of large-angle radiation from different partons in the
event.  Since the parton shower is generated with ordered opening
angles, the first gluon must be at a larger angle than the later one.
The coherence of soft large-angle radiation from the two emitters then
allows them to be described as a single emission from the
internal line imagined to be on shell, ie.~pretending that the later
emission did not happen.  Thus the earlier gluon is effectively emitted
by the lowest-order current, and both emissions should be corrected.
This is illustrated in figure~\ref{soft}.

It is worth considering what goes wrong with an algorithm that corrects
the first emission, rather than the hardest.  For events in which the
hardest emission is first there is clearly no difference, but if a soft
gluon is emitted beforehand, then no correction would be applied to the
hard gluon.  However, the probability that this occurs is strongly
(logarithmically) dependent on the infrared cutoff, leading to the
result that the hard emission cross-section of the algorithm is
unphysically dependent on its soft infrared cutoff.  That is,
\begin{eqnarray}
  \dP\hard &=& \dP\hard\me\xp{-\int_\epsilon\dP\soft\me} +
    \dP\hard\ps\left(1-\xp{-\int_\epsilon\dP\soft\me}\right) \\
  &\sim& \dP\hard\me\exp(\as\log\epsilon) +
    \dP\hard\ps\left(1-\exp(\as\log\epsilon)\right).
\end{eqnarray}
Expanding in powers of \as, this dependence might seem sub-leading,
\begin{equation}
  \dP\hard = \dP\hard\me + {\cal{O}}(\as^2),
\end{equation}
so irrelevant, but in the realm of applicability of parton showers,
logarithms of the ratio of the hard scale to the infrared cutoff are
large enough to overcome the smallness of \as, so the dependence is
formally leading order.  That is, when counting powers of \as,
$-\as\log\epsilon$ should be considered ${\cal{O}}(1)$.  Correcting only
the first emission is therefore formally inconsistent.  In our solution,
both the softer earlier emission and the later harder one would be
corrected to the matrix element, so
\begin{eqnarray}
  \dP\hard &=& \dP\hard\me\xp{-\int_\epsilon\dP\soft\me} +
    \dP\hard\me\left(1-\xp{-\int_\epsilon\dP\soft\me}\right) \\
  &=& \dP\hard\me,
\end{eqnarray}
as claimed.  This makes a particularly large difference in
angular-ordered algorithms because it is common for several soft gluons
to be emitted at large angles before reaching the hardest emission.

We finally note that the parton shower cross-section is not in general
guaranteed to be larger than the matrix-element one, so we need to be
able to enhance emission, as well as reduce it.  As we show in the
appendix, it is straightforward to uniformly enhance the emission
probability by any integer factor.  The veto algorithm can then be used
to reduce this down to the appropriate level.

\section{Summary}
We have discussed the two ways in which first-order matrix elements can
be used to improve parton shower algorithms, and showed that both must
be carefully defined to ensure self-consistency.  For the complementary
phase-space method, this requires either that the fraction of emissions
that go in to the matrix-element region be small, or that a form factor
be included to generate the hardest emission in the region exclusively,
rather than all inclusive emissions.  For the correction to the
algorithm, it requires that the correction should be applied to every
emission that is the hardest so far, rather than just to the first
emission, as was done in previous algorithms.

There is no conceptual difficulty with combining both types of
correction within the same algorithm.  Indeed for parton shower
algorithms that are capable of covering the whole of phase-space with
their first emission, the two corrections are identical for the first
emission.  The only difference is whether the phase-space points are
generated directly according to the matrix element or first according to
the parton shower algorithm, and then corrected to the matrix element.
By then going on to correct any subsequent emissions that are harder
than the first, we ensure that the correction is self-consistently
applied to the whole shower, and not just to the first emission.

Having implemented both types of correction[\ref{S1},\ref{S2}], we have
found that while the complementary phase-space method is
phenomenologically important, the correction to the algorithm has little
effect, at least within the algorithm of~[\ref{MW}], in which the parton
shower is not able to cover the whole of phase-space, but is a good
approximation in the regions it does cover.

We finally note that although such corrections have only been applied to
\ee\ or DIS so far, our arguments apply equally well to any other hard
process.  In particular, it would be straightforward to use the simple
prescription of~[\ref{S3}] to provide a matrix-element correction to the
Drell-Yan process that successfully unified the high- and low-$p_t$
regions.  This is in progress.

\section*{Acknowledgements}
I am grateful to Pino Marchesini and Bryan Webber for many discussions
of these and related subjects.  The comments of G\"osta Gustafson,
Gunnar Ingelman and Leif L\"onnblad are also gratefully acknowledged.
This work is supported in part by the EEC Programme ``Human Capital and
Mobility'', Network ``Physics at High Energy Colliders'', contract
CHRX-CT93-0357 (DG 12 COMA).

\appendix
\addtocounter{section}{1}
\section*{Appendix: The Veto Algorithm}
In this appendix we briefly recall three applications of the veto
algorithm.
\subsection{The Veto Algorithm}
Imagine that we want to generate a probability distribution
\begin{equation}
  F(x) = f(x)\xp{-\int_x^\xmax f(x)dx}\quad\mbox{with~}x<\xmax,
\end{equation}
but only know how to generate some other distribution
\begin{equation}
  G(x) = g(x)\xp{-\int_x^\xmax g(x)dx}\quad\mbox{with~}x<\xmax,
\end{equation}
with
\begin{equation}
  g(x) \ge f(x).
\end{equation}
The veto algorithm consists of the following steps to generate $F(x)$:
\begin{enumerate}
\item\label{loop}
  Generate a value of $x$ according to $G(x),$ with $x<\xmax$.
\item
  With probability $f(x)/g(x),$ keep the generated $x$ value.
\item
  Otherwise, set $\xmax=x$ and go to step~\ref{loop}.
\end{enumerate}
The probability distribution produced by this procedure satisfies the
integral equation
\begin{eqnarray}
  P(x) &=& \frac{f(x)}{g(x)}\Biggl(
    g(x)\xp{-\int_x^\xmax g(x)dx}
\nonumber\\&&
    + \bigint_x^\xmax dx'P(x')
    \frac{g(x')-f(x')}{f(x')}g(x)\xp{-\int_x^{x'}g(x)dx}
  \Biggr).
\end{eqnarray}
The term outside the bracket is the probability that the generated value
was kept, the first term in brackets is the probability that this value
was generated at the first attempt.  The second term is the integral
over all higher values, of the probability that they were generated, but
rejected, with the next value being generated at $x$.  It is
straightforward to show that this is satisfied by
\begin{equation}
  P(x) = F(x)
\end{equation}
as claimed.

\subsection{Competing Processes}\label{A2}
Imagine that we want to generate a probability distribution $F(x),$ but
only know how to generate $G(x)$ and $H(x),$ with
\begin{equation}
  f(x) = g(x)+h(x).
\end{equation}
$F(x)$ is then generated by choosing one $x$ value according to each of
$g(x)$ and $h(x),$ and using the larger.  In this case the resulting
probability distribution satisfies
\begin{equation}
  P(x) = G(x)\xp{-\int_x^\xmax h(x)dx} + H(x)\xp{-\int_x^\xmax g(x)dx},
\end{equation}
where the first term corresponds to cases where the value chosen
according to $G(x)$ is the larger, reduced by the probability that the
other value is not larger, and vice versa.  Clearly this is satisfied by
\begin{equation}
  P(x) = F(x)
\end{equation}
as claimed.  In the case that $g(x)$ and $h(x)$ correspond to different
physical processes, the process from which $x$ was generated is the one
that happened.

\subsection{Enhancing Emission}
Finally, imagine that we want to generate $F(x),$ but only know how to
generate $G(x),$ with
\begin{equation}
  f(x) = ng(x),
\end{equation}
with $n$ an integer.  This can be done as a special case of~\ref{A2}, by
considering $F(x)$ to be the sum of $n$ identical competing processes,
$G(x)$.  We then choose $n$ values of $x$ according to $G(x)$ and use
the largest of them.

\section*{References}
\def\cpc#1#2#3{Computer Phys.\ Comm.\ #1 (19#3) #2}
\def\np#1#2#3{Nucl.\ Phys.\ B#1 (19#3) #2}
\def\pl#1#2#3{Phys.\ Lett.\ #1B (19#3) #2}
\def\prep#1#2#3{Phys.\ Rep.\ #1 (19#3) #2}
\def\zp#1#2#3{Zeit.\ Phys.\ C#1 (19#3) #2}
\begin{enumerate}
\setlength{\itemsep}{0ex}
\item\label{HW}
  G.~Marchesini, B.R.~Webber, G.~Abbiendi, I.G.~Knowles, M.H.~Seymour
  and L.~Stanco, \cpc{67}{465}{92}.
\item\label{JS}
  T.~Sj\"ostrand and M.~Bengtsson, \cpc{43}{367}{87}.
\item\label{AR}
  L.~L\"onnblad, \cpc{71}{15}{92}.
\item\label{MW}
  G.~Marchesini and B.R.~Webber,
  \np{238}{1}{84}.
\item\label{Sj}
  M.~Bengtsson and T.~Sj\"ostrand,
  \pl{185}{435}{87}; \np{289}{810}{87}.
\item\label{GP}
  G.~Gustafson and U.~Pettersson,
  \np{306}{746}{88}.
\item\label{W}
  B.R.~Webber,
  \np{238}{492}{84}.
\item\label{AGIS}
  B.~Andersson, G.~Gustafson, G.~Ingelman and T.~Sj\"ostrand,
  \prep{97}{31}{83}.
\item\label{rev}
  L3 Collaboration, B.~Adeva et al.,
  \zp{55}{39}{92}. \\
  OPAL Collaboration, R.~Akers et al.,
  \zp{63}{181}{94}.
\item\label{MEPS}
  G.~Ingelman,
  in {\em Physics at HERA}, vol.~3, p.~1366.
\item\label{S1}
  M.H.~Seymour,
  \zp{56}{161}{92}.
\item\label{S2}
  M.H.~Seymour,
  Lund preprint LU TP 94-12, submitted to the International Conference
  on High Energy Physics, Glasgow, U.K., 20--27 July 1994.
\item\label{S3}
  M.H.~Seymour,
  Lund preprint LU TP 94-13, submitted to Nucl.\ Phys.\ B.
\end{enumerate}
\end{document}